\documentclass[pre,preprint,showpacs,preprintnumbers,amsmath,amssymb]{revtex4}
\usepackage{graphicx}
\usepackage{dcolumn}
\usepackage{bm}
\usepackage[mathscr]{eucal}
\usepackage{mathrsfs}

\begin{document}

\title{Interpretation of the Formation Volume of Defects in Crystal Growth}

\author{Koun Shirai}
\email{koun@sanken.osaka-u.ac.jp}
\affiliation{%
The Institute of Scientific and Industrial Research, Osaka University, 8-1 Mihogaoka, Ibaraki, Osaka 567-0047, Japan
}%

\begin{abstract}
In the calculation of the formation enthalpy of a defect, a stress term accompanying the formation volume of the defect appears. This formation volume is conventionally interpreted as the difference in volume between crystals with and without the defect. When defect formation is investigated in the study of crystal growth, the interpretation is sometimes misused by separately treating the volumes of the defect and perfect crystal. However, these two quantities are inseparable, and only the difference between them has physical reality.
\noindent

\end{abstract}


\maketitle


\section{Introduction}
\label{sec:intro}

In crystal growth, stress, whether externally applied or internally generated, has an important role in the formation of defects. It is well known that stress induces macroscopic defects, such as dislocations and stacking faults. 
Accordingly, the effect of stress on these macroscopic defects has been extensively studied for the crystal growth of silicon from a melt \cite{Mutastschiev80, Patel63,Jordan80,Yonenaga78}.
Although a similar effect of stress on the formation of microscopic point defects, namely, vacancies ($V$) and interstitials ($I$), is expected, this has not been well studied, partly because of the difficulty in detecting intrinsic point defects; normally the concentration of intrinsic point defects in silicon is of hte order of $10^{14}$ cm$^{-3}$ or less.

The recent demand for large silicon wafers is a motivating factor for better controlling defects in crystal growth. In the Czochralski (CZ) method, there is a well-known parameter for obtaining defect-free crystals, that is, ${\mathit \Gamma}=v/G$, the ratio of the pulling speed $v$ to the axial temperature gradient $G$ at the melt/solid interface, which appears in the so-called Voronkov rule \cite{Voronkov82,Voronkov99}. When ${\mathit \Gamma}$ is larger than the critical value ${\mathit \Gamma}_{c}$, vacancy-rich crystals are obtained, whereas when ${\mathit \Gamma}$ is smaller than ${\mathit \Gamma}_{c}$, interstitial-rich crystals are obtained. Therefore, it is very important to determine the value of ${\mathit \Gamma}_{c}$; currently, ${\mathit \Gamma}_{c} = 0.134$ ${\rm mm^{2}/(min \cdot K)}$ is widely used \cite{Dornberger96}. 

One of unresolved issues in crystal growth by the CZ method is whether stress affects ${\mathit \Gamma}_{c}$.
Initially, it was thought that the effect of stress is unimportant. It was estimated that the thermal stress generated in a crystal rod during growth is sufficiently small, at most of 10 MPa order, to have a sizable effect on defect formation \cite{Tanahashi00,Voronkov12}.
However, researchers soon came to recognize the effect of stress on ${\mathit \Gamma}_{c}$ \cite{Vanhellemont11,Abe11}. Since then, the debate has continued \cite{Voronkov12a,Vanhellemont12-r}.0
Sueoka, Kamiyama, and Kariyazaki (SKK) presented a theoretical study on the effect of stress, focusing on the dependence of the formation energy of intrinsic defects on pressure \cite{Sueoka12}. They concluded that compressive stress renders the crystal more vacancy-rich when the inhomogeneous distribution of stress is considered. 
Although, from the author's point of view, the SKK model is devoid of theoretical grounds for the formation volume, it was further exploited to calculate the critical value ${\mathit \Gamma}_{c}$ \cite{Sueoka13a}.
Later, an experiment was performed to examine the stress dependence of ${\mathit \Gamma}_{c}$ by Nakamura et al.~\cite{Nakamura14}. Their results showed good agreement with the theoretical prediction by the SKK model. The agreement was so impressive that their experiment seems to give compelling evidence for the SKK model.
In view of the impact on the production of silicon crystals, it is desirable to revisit the theory of SKK, which is the purpose of this short paper.
Although the study of SKK is subject to the present argument, because similar usage of the conventional interpretation is sometimes found in the community, the subject is shared with many studies in this field.



\section{Formation volume}
\label{sec:fundaments}

Let us consider an intrinsic defect $D$ ($D=I$ or $V$).
The formation enthalpy of the defect $D$ is given by
\begin{equation}
H_{f}^{D}(P) = E_{f}^{D}(P) + P v_{f}^{D},
\label{eq:form-H}
\end{equation}
under a hydrostatic pressure $P$. Let us call the second term on the right-hand side the $PV$ term. In the $PV$ term, $v_{f}^{D}$ is the formation volume of the defect $D$. Conventionally, this volume is expressed as the difference in the volume of a crystal with and without $D$,
\begin{equation}
v_{f}^{D}  = v_{N+\xi_{D}} - (v_{N} + \xi_{D} v_{1}) =
\Delta v_{\rm rx}^{D} - \xi_{D} v_{1},
\label{eq:fvol-D}
\end{equation}
where $\xi_{D}=+1$ for an interstitial and $-1$ for a vacancy. $v_{N}$ is the volume of an $N$-atom cell, whereas $v_{1}$ is the volume of the perfect crystal per atom. The difference $\Delta v_{\rm rx}^{D} =v_{N+\xi_{D}} - v_{N}$ is called the relaxation volume. Although conceptually the thermodynamic limit, $N \rightarrow \infty$, should be taken in Eq.~(\ref{eq:fvol-D}), the truncation of the crystal size is a common practice in density-functional-theory (DFT) calculations.
Equation (\ref{eq:fvol-D}) is often stated verbally,
\begin{quote}
{\bf Interpretation (I)}: Introducing an interstitial (or a vancacy) is compensated by absorbing an atom from (emitting an atom to) the surface of the crystal. 
\end{quote}
Examples of this interpretation are found in \cite{LannooBourgoin,Mehrer,Aziz97}. This interpretation is innocuous as long as hydrostatic pressure is assumed.
The problem is how to treat the inhomogeneous distribution of stress.

In the case of a distribution of stress, SKK dropped the term $P v_{1}$ from Eq.~(\ref{eq:form-H}), because they claimed that a compensated Si atom with the volume $v_{1}$ is absorbed from (or emitted to) the surface. On a free surface, the pressure vanishes, $P_{s}=0$. The $PV$ term turned to be
\begin{equation}
P \Delta v_{\rm rx}^{D} - P_{s} \xi_{D} v_{1} = P \Delta v_{\rm rx}^{D}
\label{eq:fvol-D1}
\end{equation}
With this assumption, they concluded that vacancy formation is enhanced under compressive stress because $\Delta v_{\rm rx}^{D}$ is positive. For an interstitial, the formation is suppressed. This is the main argument of SKK.

Generally, it is expected that the application of positive pressure will change a material in the direction in which the it becomes more dense. This is a consequence of the Le Chatelier principle: if an external condition of a given material is altered, the equilibrium state of the material will tend to move in a direction to oppose the change induced by the external condition. Although SKK acknowledged this principle for homogeneous stress, they arrived at the opposite conclusion for the inhomogeneous case.
The fallacy of SKK stems from unduly emphasizing the role of the surface in Interpretation (I). 
Although useful, Interpretation (I) is merely an expedient one to help to visualize the formation of defects.
The original formula of Eq.~(\ref{eq:fvol-D}) says nothing about where the compensating atom comes from or moves away. 
Each term of Eq.~(\ref{eq:fvol-D}) is an inseparable quantity, and only the net difference $v_{f}^{D}$ has physical reality. 

Let us investigate how an unphysical conclusion results when the individual terms are treated separately. Let us take the limiting case of a rigid body for defects, that is, there is no relaxation volume, $\Delta v_{\rm rx}^{D}=0$. 
Suppose that a vacancy $V$ is created on the surface of a crystal rod. Certainly, the second term on the left-hand side of Eq.~(\ref{eq:fvol-D1}), $P_{s} \xi_{D} v_{1}$, vanishes.
Next, we move the vacancy $V$ inside the rod where the pressure is $P$, leaving the emitted atom on the surface. Equation (\ref{eq:fvol-D1}) entails that only the $P \Delta v_{\rm rx}^{D}$ term corresponds to the work needed to move the vacancy there.
However, we cannot imagine that a substance with zero volume is moving in a crystal. This example shows that each term of Eq.~(\ref{eq:fvol-D}) is inseparable.
The source of the deceptive interpretation lies in the expression of Eq.~(\ref{eq:form-H}).
The volume of a defect is not a well-defined quantity, in contrast to that in the case of fluids. For solids, the $PV$ term in Eq.~(\ref{eq:form-H}) must be replaced by the energy change due to the elastic deformation $V \sum_{ij} \sigma_{ij} \epsilon_{ij}$, namely, the product of stress ($\sigma_{ij}$) and strain ($\epsilon_{ij}$) tensors. The elastic deformation energy has a nonlocal character, and thereby the decomposed terms in Eq.~(\ref{eq:fvol-D}) have meaning only when they are considered as a set.

With this understanding of the formation volume, let us consider a correct formalism for the deformation energy in an inhomogeneous distribution of stress. On the assumption that the local equilibrium is valid, the enthalpy $H_{f}^{D}({\mathbf x})$ can be defined at each point ${\mathbf x}$ of a crystal. Using the position-dependent pressure $P({\mathbf x})$, $H_{f}^{D}({\mathbf x})$ can be expressed by
\begin{equation}
H_{f}^{D}({\mathbf x}) = E_{f}^{D} + P({\mathbf x}) v_{f}^{D}({\mathbf x}),
\label{eq:x-dep-H}
\end{equation}
provided that $E_{f}^{D}$ is independent of $P$. 
In equilibrium, the distribution of $D$, $N_{D}({\mathbf x})$, becomes $N_{D}({\mathbf x}) = N_{0} \exp( -H_{f}^{D}({\mathbf x})/k_{\rm B}T)$, where $k_{\rm B}$ is Boltzmann's constant. For an interstitial, $v_{f}^{I}$ is negative, from which it follows that interstitials tend to move to the region under compression.
Similarly, vacancies tend to move to the region under expansion.
By defining the chemical potential as $\mu = k_{\rm B}T \ln (N_{D}({\mathbf x})/N_{0})$, we have the equation
\begin{equation}
\nabla \mu_{D}({\mathbf x}) = - \nabla H_{f}^{D}({\mathbf x}).
\label{eq:grad-mu}
\end{equation}
This equation states that the stress acting on defects is balanced with the diffusion force of the defects, $-\nabla \mu_{D}({\mathbf x})$, at which thermodynamic equilibrium is established.

\section{Discussion}
The effect of stress on the formation volume was implemented by SSK in the expression of the critical ratio $\Gamma_{c}$,
\begin{equation}
{\mathit \Gamma}_{c} = \frac{C_{I}D_{I} H_{f}^{I} -C_{V}D_{V} H_{f}^{V}}{ 
  (C_{V} - C_{I}) k_{\rm B} T^{2}},
\label{eq:modifiedG0}
\end{equation}
through the pressure dependence of the formation enthalpy $H_{f}^{D}$. Here, $C_{D}$ and $D_{D}$ are the equilibrium concentration and diffusivity of the defect species $D$, respectively. The values of these quantities are those at the melting point. 
Nakamura {\it et al.}~examined the application of the SSK model to crystal growth by the CZ method \cite{Nakamura14}. In a grown crystal rod, there is a distribution of $v$ and $G$ in the axial ($z$) and radial ($r$) directions. After identifying the boundary of the $V$-rich and $I$-rich regions, they sampled $v/G$ on the boundary. In this manner, they obtained the stress dependence of ${\mathit \Gamma}_{c}$. However, the original model of Voronkov was a one-dimensional model in the $z$-direction \cite{Voronkov82,Voronkov99}. Taking the radial distribution into account may introduce further complications. Presently, it is better to reserve judgement on whether their experiment supports the SKK model.

In experiments, the high-pressure doping method has been utilized for a wide range of materials, for example, B-doped diamond \cite{Ekimov04a}, Ba-doped Si clathrates \cite{Yamanaka00}, and Fe-based layered compounds \cite{Shirage08}, although the merits of the method are not fully understood.
For the diffusion of impurities, the effect of pressure in very high pressures of about 1 GPa was observed by Zhao {\it et al.}~\cite{Zhao99}. At such a high pressure, it is reasonable to observe a sizable effect on the formation energy. For the high-pressure preparation of boron crystal at pressures in a similar range, the variation in the concentration of interstitial atoms was experimentally observed. The variation was well accounted for by the pressure dependence of the formation volume \cite{Uemura16}. The formation volume of a point defect in boron is similar to that in silicon, 10 to 50 meV/GPa. Hence, at pressures higher than 1 GPa, it is reasonable to observe the effect of pressure on the formation of defects. In contrast to that observed in contemporary high-pressure experiments, the pressure in the CZ method is normally less than 0.5 GPa, and the effect of pressure is small. However, in the present case of intrinsic defects, the issue is the effect on defects with very low concentrations of the order of $10^{15}$ cm$^{-3}$. Vanhellemont pointed out that only the difference in $H_{f}^{D}$ between $V$ and $I$ is relevant in determining ${\mathit \Gamma}_{c}$ \cite{Vanhellemont12-r}. Hence, there seems to be no reason to exclude the effect of stress on ${\mathit \Gamma}_{c}$.


\section{Conclusion} 
\label{sec:conclusion}
In the conventional expression of the formation volume given by the difference in volume between crystals with and without the defect, the two volumes are inseparable, and only the difference between them has physical reality.

\begin{acknowledgments}
The author thanks N.~Inoue for attracting the author's attention to this problem. He also thanks T. Abe and T. Takahashi for their discussion on crystal growth.
\end{acknowledgments}


\end{document}